\documentstyle[aps,prl]{revtex}
\begin{document}
\draft
\twocolumn[
\title{Driven Maps and the Emergence of Ordered Collective Behavior\\
in Globally Coupled Maps}
\author{A. Parravano and M.G. Cosenza}
\address{\hspace{-0.3cm}Centro de Astrof\'{\i}sica Te\'orica, Facultad de Ciencias,
Universidad de Los Andes, A.P. 26 La Hechicera, M\'erida 5251,
Venezuela.} 
\date{Submitted to Phys. Rev. Lett.}
\maketitle
\vspace{-1.0cm}
\begin{abstract}
\widetext
A method to  predict the emergence of different kinds
of ordered collective behaviors 
in systems of globally coupled chaotic maps is proposed.  
The method is based on the analogy between globally coupled maps
and a map subjected to
an external drive. A vector field which results from this analogy 
appears to govern the transient evolution of the globally
coupled system. General forms of
global couplings are considered.  Some simple applications 
are given. 
 
\narrowtext
\end{abstract}
]

Coupled map lattices constitute useful models for the study of 
spatiotemporal processes in a variety of contexts \cite{Survey}.
There has been recent interest in the investigation of
the emergence of ordered collective behaviors (OCB) 
in systems of
interacting chaotic elements by using coupled map lattices 
\cite{Kan1,Gupte,ChateA,Chate1}. 
Such cooperative phenomena have been considered relevant in 
many physical and biological
situations \cite{Roy,Marcus,Kuramoto,Kan2}.  
In particular, globally coupled maps (GCM) \cite{KanGCM} can exhibit OCB such as: 
a) formation of clusters, i.e., differentiated subsets
of synchronized elements within the network \cite{Kan3};
b) non-statistical
properties in the fluctuations of the mean field of the ensemble
\cite{Kan3,Perez}; c) global quasiperiodic motion \cite{Kan4,Pikovsky}; and
different collective phases depending on the parameters of the system
\cite{Kan4}.

Much effort has been dedicated to establishing the necessary conditions for
the emergence of various types of OCB,
mostly involving direct numerical simulations on the whole 
globally coupled system. However, this 
direct procedure gives little information
about the mechanism for the emergence of OCB. In this letter, 
we propose an alternative 
method which allows us to gain insight on the conditions for the emergence  
of  specified types of collective behavior.  
 
The procedure is suggested by the analogy between a globally coupled map system
\begin{equation}
\label{ec1}
x_{t+1}^n=(1-\epsilon) f(x_t^n) + \epsilon H(x_t^1,x_t^2, \ldots , x_t^N)\, ,
\end{equation}
and an associated driven map 
\begin{equation}
\label{ec2}
s_{t+1}=(1-\epsilon) f(s_t) + \epsilon L_t\, .
\end{equation}

In eq.~(\ref{ec1}), $x_t^n$ ($n=1,\ldots,N$) 
is the state of the $n$th element at 
discrete time $t$; $N$ is the size of the system; $\epsilon$ is 
the coupling parameter;
and $f(x)$
describes the local dynamics. A commonly used form for $H$ is
the mean field coupling \cite{KanGCM,Perez,Pikovsky}, however the global coupling 
function $H _t=H(x_t^1,x_t^2, \ldots , x_t^N)$ in eq.~(\ref{ec1}) is
assumed to be any function invariant to argument permutations. 
In eq.~(\ref{ec2}),
$s_t$ is the state of the driven map at discrete time $t$, $f(s_t)$ 
is the same local dynamics
than in eq.~(\ref{ec1}), and $L_t$ is the driving term.
In general, $L_t$ may be any function of time.

Here we focus on the relationship between periodically driven maps
and the emergence of periodic and quasiperiodic OCB in GCM. In this case, the 
asymptotic collective
behavior is characterized by periodic or quasiperiodic
time evolution of the mean state $\langle x \rangle_t$ and of
the coupling function $H_t$ for $t > t_a$, where $t_a$ is the
transient time required to reach the OCB.
The elements of the system  follow a similar behavior to 
$H_t$, and in general are segregated in 
clusters which tend to be out of phase \cite{KanGCM}.

In the case of an OCB consisting of 
$K$ clusters with period $P$, the asymptotic behavior of the system 
can be characterized by a $K \times P$ matrix 
\begin{equation}
\label{fi}
{\mbox{\boldmath$\chi$}}=
\left(
\begin{array}{ccc}
\chi _1^1 & \ldots & \chi _1^K \\
\vdots & \ddots & \vdots \\
\chi _P^1 & \ldots & \chi _P^K \\
\end{array}
\right) \; ,
\end{equation}
where the $k$th column contains
the temporal sequence of the $P$ values adopted by the elements belonging 
to the $k$th cluster;
and the $i$th row displays the state of all clusters at time $i$. 
The time step $t$ have been replaced by the 
index $i$ which runs from 1 to the periodicity $P$. 
Therefore, the $P$ asymptotic values $\Phi_i$ adopted by the 
global coupling 
$H_{t>t_a}$ 
are
\begin{equation}
\Phi_i=H(\underbrace{\chi_i^1,\ldots,\chi_i^1}_{N_1 \, \mbox{times}}
,\ldots,\underbrace{\chi_i^k,\ldots,\chi_i^k}_{N_k \, \mbox{times}}
,\ldots,\underbrace{\chi_i^K,\ldots,\chi_i^K}_{N_K \, \mbox{times}}),
\end{equation}
where $N_k$ is the number of elements in the $k$th cluster and 
$\sum_k N_k = N$.
Thus, we may express the asymptotic temporal behavior of  $H_{t>t_a}$
by  
the vector $\vec{\Phi}=(\Phi_1,\ldots,\Phi_i,\ldots,\Phi_P)$.

When the driven map, eq.~(\ref{ec2}), is subjected
to a periodic drive $L_t$ with period $P$, 
the long-term response of $s_t$ depends on the initial condition $s_1$, and on
the specific sequence of values
adopted by $L_t$, which we will denote as the vector 
$\vec{L}=(L_1,\ldots,L_i,\ldots,L_P)$.
The response $s_t(\vec{L},s_1)$ may be periodic on 
a region $R$ of the $P$-dimensional space  
spanned by all possible vectors $\vec{L}$. 
For a given $\vec{L} \in R$,  
there are, usually, $J$ different asymptotic periodic responses of
$s_t$ with period $M$,
each one associated to a set 
of initial conditions $\{s_1\}_j$, where $j=1,2,\ldots,J$.
If we denote by $\sigma _i^j$ the asymptotic periodic response 
of $s_t(\vec{L},s_1)$ that is reached for $s_1 \in \{s_1\}_j$,
then the possible
asymptotic responses can be represented by the $J \times M$ matrix 
\begin{equation}
{\mbox{\boldmath$\sigma$}}=
\left(
\begin{array}{ccc}
\sigma _1^1 & \ldots & \sigma _1^J \\
\vdots & \ddots  & \vdots \\
\sigma _M^1 & \ldots & \sigma _M^J \\
\end{array}
\right) \; ,
\end{equation}
where the $j$th column contains the asymptotic periodic response of $s_t$
for the initial states  $s_1 \in \{s_1\}_j$; and the $i$th row 
displays the state of all asymptotic responses at time $i$.
As before, the time step $t$ has been replaced by the index $i$ 
running from 1 to the
periodicity $M$ of the response. Obviously, this scheme is valid
if all the asymptotic responses ($j=1,\ldots,J$) have the same 
periodicity $M$
(not necessarily $M=P$).
The region $R$ may consist of various subregions $R_M^J$, where 
$J$ different asymptotic responses of periodicity $M$ occur.
In particular, the subregion $R_P^J$ is relevant
since there the asymptotic responses of the driven map have 
the same periodicity 
than the drive, and consequently the analogy between eq~(\ref{ec1})
and eq.~(\ref{ec2}) can be established. Note that if $\vec{L}=\vec{\Phi}$,
the columns of the OCB matrix ${\mbox{\boldmath$\chi$}}$ are contained in
the response matrix~${\mbox{\boldmath$\sigma$}}$. 
\setcounter{figure}{0}
\renewcommand{\thefigure}{1}
\begin{figure}[htb]
\vspace{7.8cm}
\includegraphics{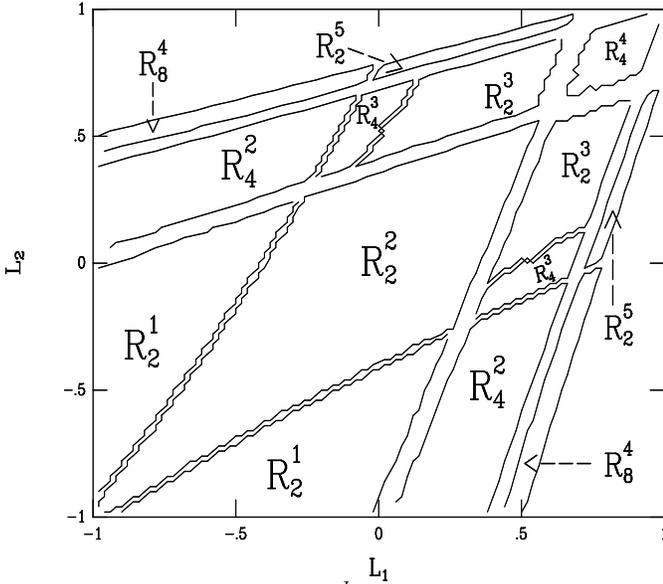}
\caption[]{Main subregions $R_M^J$ for the driven logistic map with $P=2$.
Parameter values are $r=1.7$ and $\epsilon=0.2$.}
\end{figure}
As an illustration,
figure~1 shows the main subregions $R_M^J$ resulting from eq.~(\ref{ec2})
for a 
logistic map $f(x)=1-r x^2$, driven by  different $L_t$ 
with the same period $P=2$.  
Parameter values are $r=1.7$ and
$\epsilon=0.2$. 
The subregions $R_M^J$ indicate where
$J$ asymptotic responses of periodicity $M$ occur when the components
$L_1$ and $L_2$ of $\vec{L}$ lie in the range
$[-1,1]$. The structure of this diagram is actually more complex; there are
small zones in between  
the marked subregions where the driven map reaches a variety of
periodic responses,  
(not shown in fig.~1).
We have also observed shrimp-like structures for $L_1$ and $L_2$
outside the range $[-1,1]$, similar to those reported for two-parameter
maps \cite{Gallas}. 

Now consider
the evolution of the GCM towards an OCB of period $P$.
During this transient time, the elements of the system 
segregate in ``swarms" which progressively shrink, and 
eventually form clusters.
Simultaneously, $H_t$ evolves towards its asymptotic
periodic behavior $\vec{\Phi}$ given by eq.~(\ref{fi}). 
The convergence of $H_t$ towards $\vec{\Phi}$ can be 
represented in the same P-dimensional space spanned by the vectors $\vec{L}$
as a trajectory joining the point $(H_1,\ldots,H_P)$ to 
$(H_{P+1},\ldots,H_{2P})$
to $(H_{2P+1},\ldots,H_{3P})$  $\ldots$ to 
$(\Phi_1,\ldots,\Phi_{P})=\vec{\Phi}$.

Suppose that we want to find out if, for a given coupling function
$H_t$ in eq.(\ref{ec1}), an OCB with period $P$ and $K$ 
clusters can be observed. The distribution
of elements in clusters corresponds to a partition $\{p_1,p_2,\ldots,p_K\}$, 
where $p_k$ is the fraction of elements in the $k$th cluster.
Then, based on the analogies presented above,
the following procedure may be employed:

a) Determine the region $R_P^K$ in which the response matrix 
${\mbox{\boldmath$\sigma$}}$ is $K \times P$, by
exploring  
the asymptotic responses
of the associated driven map 
for different drives $L_t$ of the same period $P$ in eq.~(\ref{ec2}).

b) Construct the associated coupling vector
$\vec{\Theta}=(\Theta_1, \ldots, \Theta_i, \ldots, \Theta_P)$,
whose components are given by
\begin{equation}
\Theta_i=
H(\underbrace{\sigma_i^1,\ldots,\sigma_i^1}_{N \, p_1  \, \mbox{times}}
,\ldots,\underbrace{\sigma_i^k,\ldots,\sigma_i^k}_{N \, p_k \, \mbox{times}}
,\ldots,\underbrace{\sigma_i^K,\ldots,\sigma_i^K}_{N \, p_K \, \mbox{times}})
\end{equation}

c) In the space $R_P^K$ construct the vector field
\begin{equation}
\vec{V}
=\vec{L}-\vec{\Theta}\, .
\end{equation}

If in the region $R_P^K$ the vector field $\vec{V}$ is convergent 
towards the locus where 
$\vec{V}=0$, then the chosen OCB can take place for appropriate
initial conditions in the GCM system. 

As shown in the examples bellow, the vector field $\vec{V}$ acts as an 
indicator  
of the transient trajectory of $H_t$; that is, the region of convergence of
$\vec{V}$ towards  $\vec{V}=0$ is
related to the basin of attraction of the chosen OCB in the space of the $P$ 
initial values $(H_1,\ldots,H_P)$.
The field $\vec{V}$ is particulary efficient in pointing the transient 
trajectory of $H_t$
when the associated driven map reaches its asymptotic periodic response
in few
iterations. 
Notice that the vector field $\vec{V}$ can be calculated without
previous knowledge of the existence of chosen OCB in the corresponding GCM. 
Moreover,
the first step of the procedure (and the most time consuming) 
must be executed 
only once; 
thereafter, steps (b) and (c) can be performed repeatly for many
different 
coupling functions $H_t$ and different partitions $\{p_1,p_2,\ldots,p_K\}$.

In order to show applications of the procedure, we choose
globally coupled logistic maps
$f(x)=1-r x^2$, and look for an OCB consisting of two 
clusters of period two. Parameter values are fixed at $r=1.7$ and
$\epsilon=0.2$ in what follows. 
\setcounter{figure}{0}
\renewcommand{\thefigure}{2}
\begin{figure}[htb]
\vspace{7.8cm}
\includegraphics{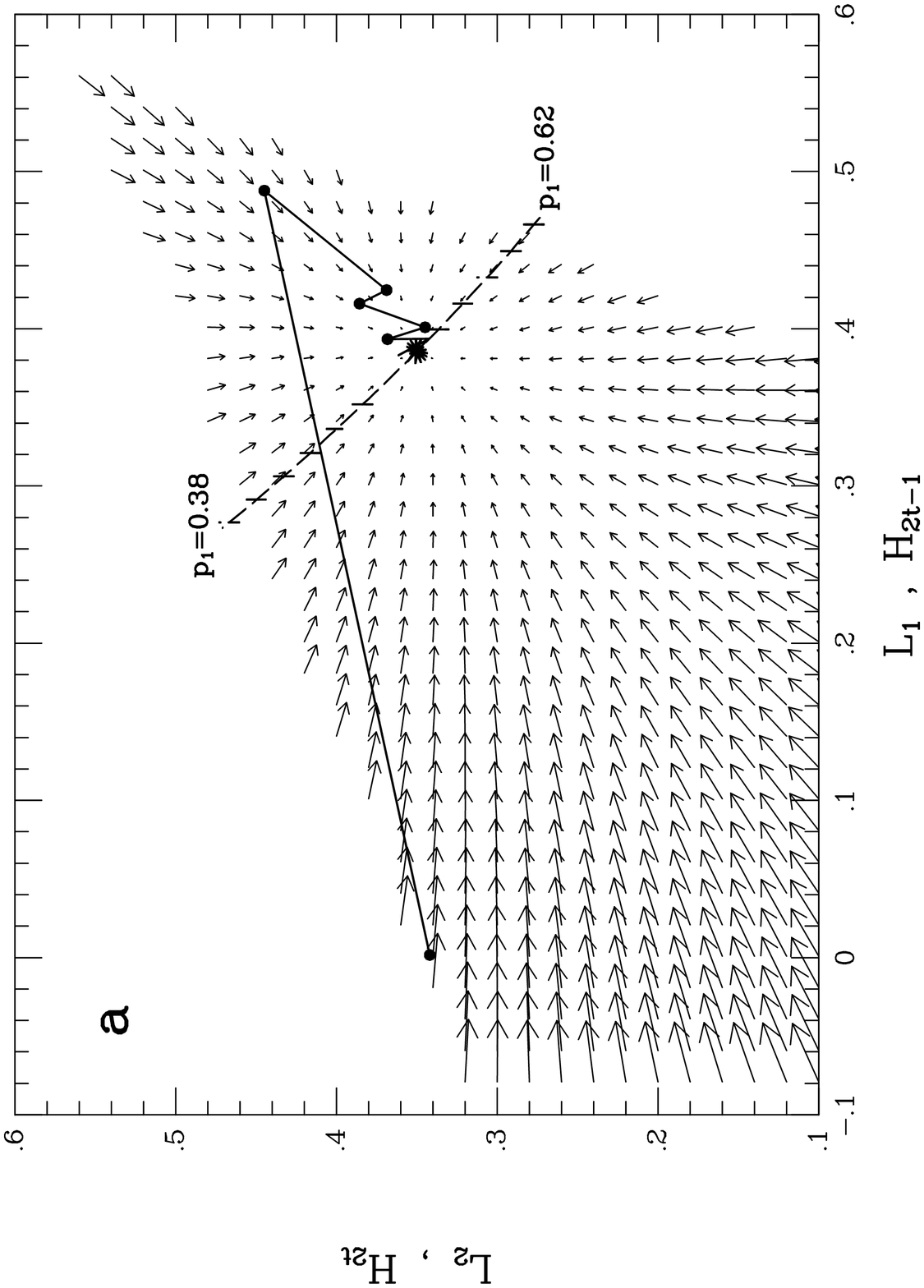}
\end{figure}
\vspace{-0.5cm}
\setcounter{figure}{0}
\renewcommand{\thefigure}{2}
\begin{figure}[htb]
\vspace{7.8cm}
\includegraphics{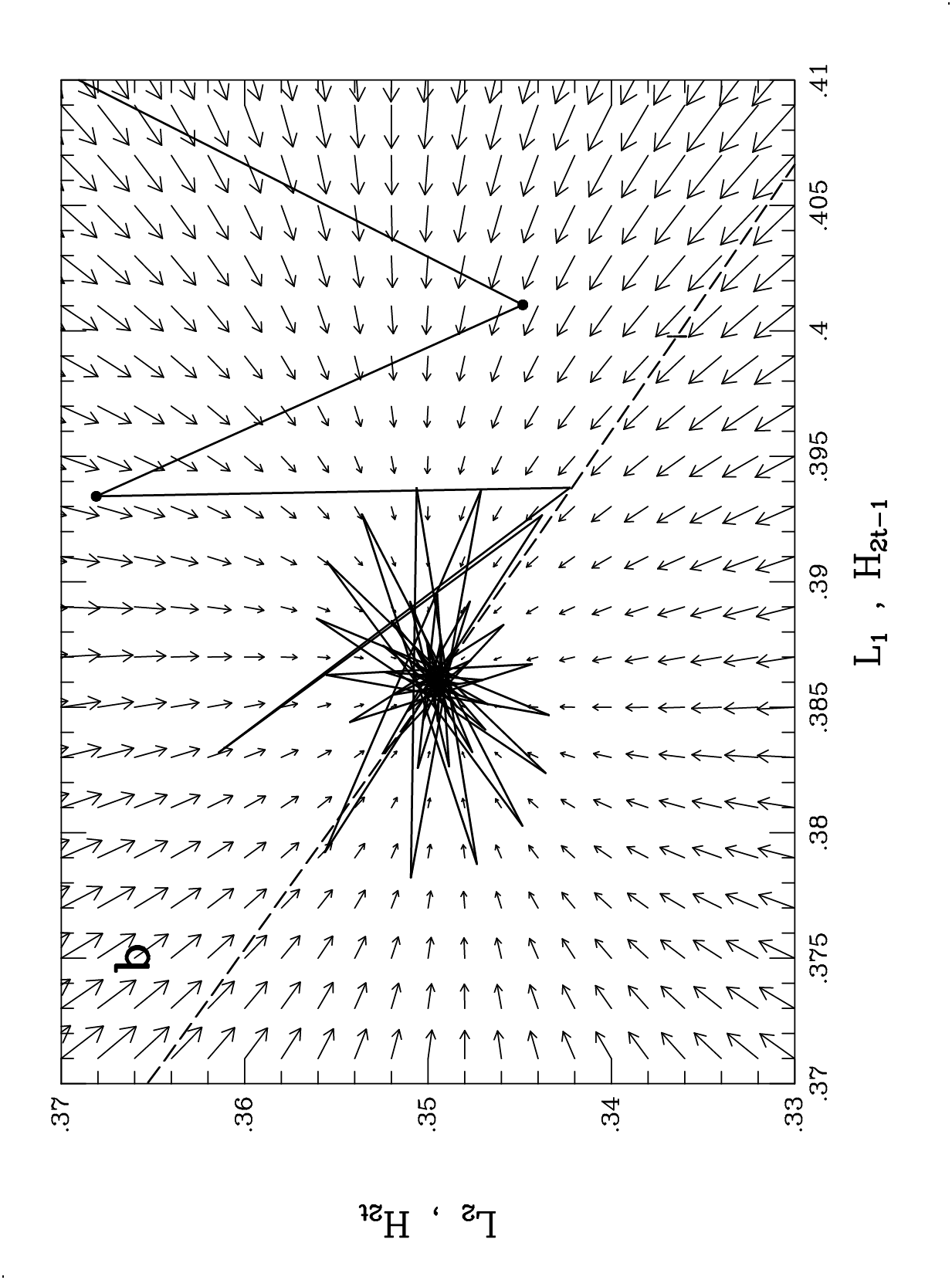}
\caption[]{a) The vector field $\vec {V}$ and the trajectory of $H_t$
for the arithmetic mean global coupling. b) Magnification of (a).
}
\end{figure}
As a first example, the coupling function $H_t$ is assumed to be the
arithmetic mean; i.e.,
$H_t=\frac{1}{N} \sum_{n=1}^{N} x_t^n = \langle x \rangle_t $.
Therefore, 
$\Theta_{i}=p_1 \sigma_i^1 + p_2 \sigma_i^2$, 
and 
$ V_i=\Theta_i-L_i$, $(i=1,2)$. Figure~2(a) shows the
vector field $\vec{V}(L_1,L_2,p_1)$ in the region $R_2^2$ 
for a partition $p_1=0.477$ and $p_2=1-p_1$.
The vector
field $\vec {V}(L_1,L_2,p_1)$ is plotted as arrows of length proportional 
to $|\vec {V}|$,
direction given by $\tan^{-1} (V_1/V_2)$, and origin at
$(L_1,L_2)$.   
Figure~2(a) also shows the superposition of the trajectory of the coupling
function $H_t$
constructed 
by joining $(H_1,H_2)$ to $(H_3,H_4)$ $\ldots$ to $(\Phi_1,\Phi_2)$. 
Random initial conditions of $N=2000$ elements uniformly distributed 
on $[-1,1]$ are used, i.e., $\langle x \rangle_1=0$.  
For these initial conditions, the GCM rapidly collapses
into two clusters with
the chosen partition, i.e.,
$N_1=954$ and $N_2=1046$, respectively. 
For other initial conditions, different partitions result. In those cases 
the vector field $\vec{V}$ maintains the same
appearance, except that the point where $\vec{V}=0$ moves along
the dashed curve in fig.~2(a) as $p_1$ varies. 
The two labeled values of $p_1$ at the ends of the dashed curve in fig.~2(a)
indicate 
the
position of the convergent point $\vec{V}=0$ for these critical values of $p_1$. 
An OCB of period two with two clusters can not be observed for values of 
$p_1$
outside this interval. 
Note that for the first few iterations, the $H_t$ 
transient trajectory
is roughly suggested by the field $\vec {V}$, since at these early
stages the cluster are being formed.
As time progresses, the vector field $\vec {V}$ becomes a better
indicator of the evolution of the GCM system toward its OCB, as showed in 
fig.~(2b).

In a second example, we consider the geometric mean of the moduli
of the values of the elements of the
system as the coupling function, i.e., 
$H_{t}= \prod_{n=1}^{N} \left|x_t^n \right|^{\frac{1}{N}}$.
Therefore, in the region $R^2_2$, 
$\Theta_{i}= |\sigma_i^1|^{p_1} \times  |\sigma_i^2|^{1-p_1}$, $(i=1,2)$. 
Figure~3 again shows the corresponding field $\vec{V}$ together with
the trajectory of the coupling function $H_t$. The same initial conditions and
system size as in the first example have been used.
In this case, the convergence to the OCB is faster in comparison to the 
first example. 

\setcounter{figure}{0}
\renewcommand{\thefigure}{3}
\begin{figure}[htb]
\vspace{7.8cm}
\includegraphics{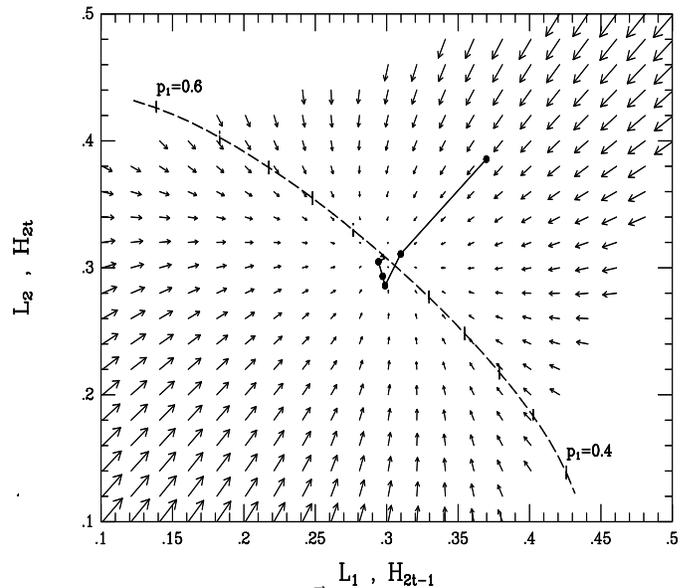}
\caption[]{The vector field $\vec {V}$ and the trajectory of $H_t$ 
for the geometric mean of the moduli coupling.
}
\end{figure}

Finally, the method can be used to infer a global coupling function
capable of producing a quasiperiodic OCB with two clusters. 
This can be achieved by modifying  the coupling function
of a known case of periodic OCB, just near its convergent point 
$\vec{V}=0$. 
Then, we may expect 
that the modified GCM will not reach its original periodic OCB, but 
would remain 
close to it.
For the first example, this can be done by noticing first 
that when  $\vec {L}$ is near the convergent 
point, one of the  two asymptotic responses of the driven map 
adopts a value close to $0.86$ every two time steps. 
Then, the coupling function of the first example
may be modified in such a way
that only near $0.86$ the function is drastically affected.
\setcounter{figure}{0}
\renewcommand{\thefigure}{4}
\begin{figure}[htb]
\vspace{7.8cm}
\includegraphics{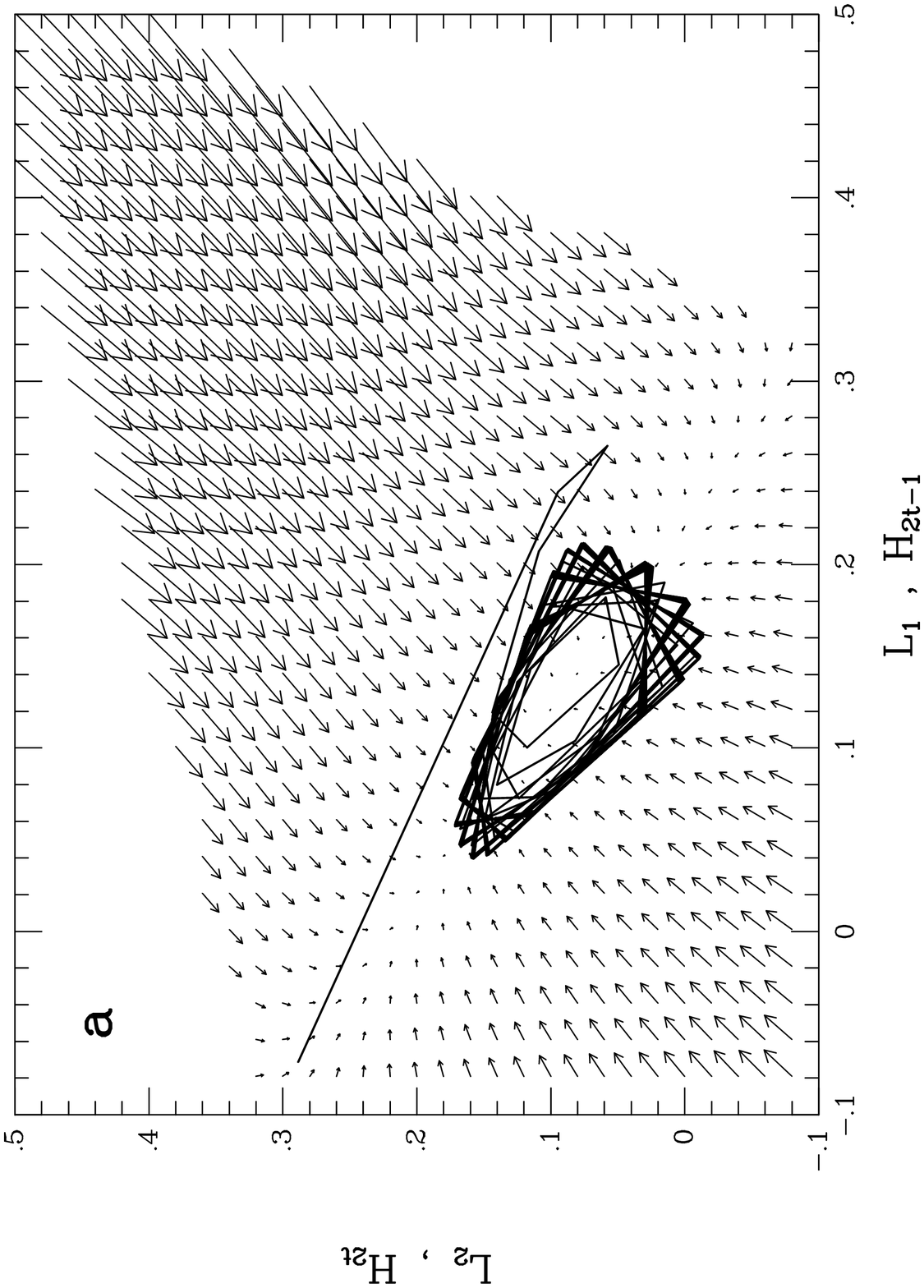}
\end{figure}
\vspace{-0.5cm}
\setcounter{figure}{0}
\renewcommand{\thefigure}{4}
\begin{figure}[htb]
\vspace{7.8cm}
\includegraphics{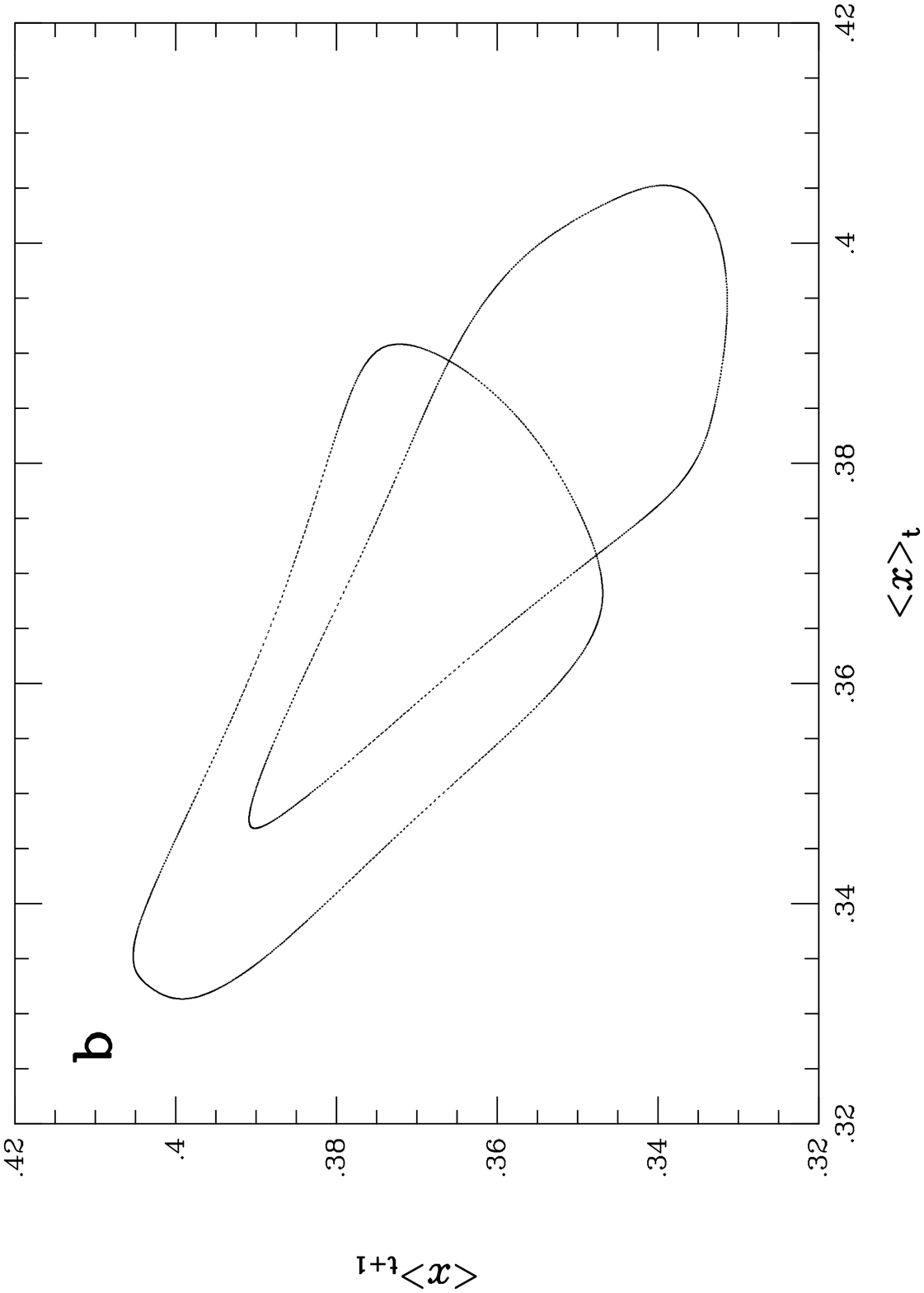}
\caption[]{a) The vector field $\vec {V}$ and the trajectory of $H_t$
for the modified arithmetic mean global coupling.
b) The corresponding asymptotic return map of the mean state of the system.
}
\end{figure}  
We have tried with coupling functions of the type:
\begin{equation}
\label{modi}
H_{t}=\frac{1}{N} \sum_{n=1}^{N} x_t^n \left[1-a \left(1-\exp \left(\frac{1}
{b(x_t^n-0.86)^2} \right) \right) \right]
\, . 
\end{equation}
The results for parameters values $a=1.7$ and $b=10^3$ in eq.~(\ref{modi})
are shown in figure~4(a). Additionally, 
the return map of the arithmetic mean $<x>_{t+1}$ vs. $<x>_{t}$ is 
shown in fig.~4(b).
This example shows the usefulness of the method for 
designing globally coupled systems
with specific features.
Quasiperiodic OCB appears associated to the type of
convergence of the field $\vec{V}$ occuring in this example.
This type of convergence can occur for other $f(x)$ and $H_t$.

A variety of clustered OCB can be
predicted for coupled logistic maps with $r_{\infty} < r < 2$, provided
that the value of $\epsilon$ is in the appropriated range,
in accordance with fig.~3 of \cite{KanGCM}.
In the above examples, the field $\vec{V}$ can be
represented on a plane, but for $P > 3$,
projections of $\vec{V}$ can reveal its global
structure. As $P$ increases, the computation time
increases potentially with $P$. Then, numerical methods to search
for convergence towards $\vec{V}=0$ may be used to speed the process.

In summary, we have presented a method to study 
the emergence of ordered collective behaviors in globally coupled maps, 
based on the analogy of these systems with a driven map. 
A vector field defined on the space of the drive term in the driven map
acts as an indicator of the evolution of the coupling function in the GCM.  
The method can be used to predict if specific types of OCB can take place
in a GCM system and to visualize its associated basin of attraction.  
The limitation of this method lies in the fact that the driven
map (eq.~(\ref{ec2})) must posses periodic asymptotic responses
so that the matrix~${\mbox{\boldmath$\sigma$}}$ can be defined.
A large family of maps fulfill this condition.
The examples presented here show that progress
in the understanding of the collective behaviors of GCM can be made by 
investigating its relation with driven maps. Extensions of this method could be
applied to phenomena such as
control of chaos,
chaotic synchronization, phase segregations, and intermittent OCB.

\section*{Acknowledgments}
This work was supported by C.D.C.H.T.,
Universidad de Los Andes, M\'erida, Venezuela.

\end{document}